# A new double-layered kagome antiferromagnet ScFe$_6$Ge$_4$

Mohamed Abdelkareem Kassem[a,1], Taiki Shiotani[a,*], Hiroto Ohta[b], Yoshikazu Tabata[a], Takeshi Waki[a], Hiroyuki Nakamura[a]

[a] Department of Materials Science and Engineering, Kyoto University, Kyoto, 606-8501, Japan

[b] Department of Molecular Chemistry and Biochemistry, Doshisha University, Kyotanabe, 610-0321, Japan

*Corresponding author.

Email address: shiotani.taiki.48e@st.kyoto-u.ac.jp (T. Shiotani)

[1]Permanent address: Department of Physics, Faculty of Science, Assiut University, 71516 Assiut, Egypt

**Abstract**

ScFe$_6$Ge$_4$ with the LiFe$_6$Ge$_4$-type structure (space group $R\bar{3}m$), which has a double-layered kagome lattice (18$h$ site) of Fe crystallographically equivalent to that of a well-known topological ferromagnet Fe$_3$Sn$_2$, is newly found to be antiferromagnetic (AFM) with a high Néel temperature of $T_\mathrm{N} \approx 650$ K, in contrast to the ferromagnetic (FM) ground state previously proposed in a literature. $^{45}$Sc nuclear magnetic resonance experiment revealed the absence of a hyperfine field at the Sc site, providing microscopic evidence for the AFM state and indicating AFM coupling between the bilayer kagome blocks. The stability of the AFM structure under the assumption of FM intra-bilayer coupling is verified by DFT calculations.

**Keywords**: A. ScFe$_6$Ge$_4$; C. LiFe$_6$Ge$_4$-type; C. double-layered kagome lattice; E. NMR

Recently, a number of non-trivial quantum phases have been discovered in magnetic materials consisting of the kagome lattice [1]. One of the intensively studied materials is the ferromagnetic (FM) $Fe_3Sn_2$ with a double-layered kagome lattice of Fe [2], which is a well-known topological ferromagnet (see, e.g., [3]). The space group of $Fe_3Sn_2$ is $R\bar{3}m$ and Fe occupies a site with Wykoff symbol 18$h$ [2]. The Fe site is characterized by two adjacent breathing kagome layers (bilayer block). As a family of materials with the Fe sublattice crystallographically equivalent to that in $Fe_3Sn_2$, $RFe_6Ge_4$ compounds ($R$ = Li, Mg, Zr, Sc) with the $LiFe_6Ge_4$-type structure are known [4] (compare Figs. 1(a) and (b)). However, their magnetic properties are poorly understood. On the other hand, the magnetic properties of the $RM_6X_6$ family ($R$ = Gd–Tm, Mg, Sc, Y, Lu, Ti, Zr, Hf, Nb; $M$ = Fe, Mn; $X$ = Ge, Sn), with a local structure similar to $RFe_6Ge_4$, have been studied intensively [5-7] and their topological nature is also attracting attention recently (see, e.g., [8, 9]).

In $RFe_6Ge_4$, the Fe bilayer blocks and $RGe_8$ hexagonal bipyramidal layers ($R$ is the center of the Ge dodecahedron) are alternately stacked in the $c$-direction (Fig. 1(b)) [4]. The $Sn_8$ bipyramidal layer in $Fe_3Sn_2$ is replaced by the $RGe_8$ layer in In $RFe_6Ge_4$. The vertices of the $Sn_8$ bipyramids in $Fe_3Sn_2$ do not penetrate the Fe kagome plane, while those of the $ScGe_8$ in $RFe_6Ge_4$ slightly penetrate the kagome plane (see Figs. 1(a) and (b)). On the other hand, $RFe_6Ge_6$ of the $HfFe_6Ge_6$-type ($P6/mmm$) has Fe monolayers (non-breathing kagome, 6$i$ site) stacked in phase, with the $RGe_8$ bipyramidal layers and the Ge honeycomb layers alternating between the Fe kagome layers (Fig. 1(c)) [10, 11]. $RFe_6Ge_6$ is antiferromagnetic (AFM) when $R$ is a nonmagnetic element [12-15], while $Fe_3Sn_2$ is FM [16]. The magnetic anisotropy of both $Fe_3Sn_2$ and $RFe_6Ge_6$ is basically uniaxial with the $c$-axis as the easy axis: a canted component appears at low temperatures, but both show collinear structures just below the Curie and Néel temperatures ($T_C$ and $T_N$), respectively [12, 15-17]. The Fe moment of $Fe_3Sn_2$ is ~2.0 $\mu_B$ [16], while that of $RFe_6Ge_6$ is 1.5–2.1 $\mu_B$ [13, 15], depending on $R$. These three crystal structures have in common that they are composed of the Fe kagome lattice, but the spacing and phase between the Fe layers are different, resulting in a diversity of magnetic coupling between the layers.

The only experimental study of the magnetic properties of the $RFe_6Ge_4$ series is for $ScFe_6Ge_4$ [18]. From magnetization measurements of polycrystalline $ScFe_6Ge_4$, the authors claimed that $ScFe_6Ge_4$ is FM with $T_C$ = 491 K and a spontaneous magnetization $M_s$ of ~0.5 $\mu_B$ per formula unit (at 3 K), i.e. ~0.1 $\mu_B$/Fe. Mössbauer spectroscopy revealed an internal magnetic field at the Fe site $|H_{hf}(Fe)|$ of 191 kOe (common at room temperature (RT) and 5 K). Assuming a standard hyperfine coupling of for 3d transition metal elements (~ −100 kOe/$\mu_B$), the magnitude of the Fe moment $\mu_{Fe}$ is estimated to be ~1–2 $\mu_B$. This value is an order of magnitude larger than $M_s$. In other words, $M_s$ is unusually small, making it unlikely that ScFe$_6$Ge$_4$ is in a simple ferromagnet. Furthermore, they argued from DFT calculations that the spin-polarized state is stable, but at the same time they stated that the FM state is more stable than AFM states. Assuming a FM structure, they estimated $\mu_{Fe} \approx 1.8$ $\mu_B$. This is in agreement with $|H_{hf}(Fe)|$ observed by Mössbauer spectroscopy, but not with $M_s$.

In this study, we newly found that ScFe$_6$Ge$_4$ is not FM but AFM with $T_N \approx 650$ K. $^{45}$Sc nuclear magnetic resonance (NMR) shows that the internal field at the Sc site $H_{hf}(Sc)$ is canceled in its magnetically ordered state, which microscopically proves that ScFe$_6$Ge$_4$ is not FM and indicates that the magnetic coupling between the Fe bilayer blocks is AFM. We also show by DFT calculations that the AFM state is more stable than the FM state.

We have synthesized ScFe$_6$Ge$_4$ from Sc (Johnson Matthey, 99.9% purity), Fe (Johnson Matthey, 99.95%), and Ge (Rare Metallic, 99.999%) by arc melting in an argon atmosphere. Powder X-ray diffraction measurements at RT showed that the sample has the LiFe$_6$Ge$_4$-type structure with lattice constants $a = 5.071$ Å and $c = 20.053$ Å, which are in good agreement with the literature values $a = 5.066$ Å and $c = 20.013$ Å [4], $a = 5.079$ Å and $c = 20.009$ Å [18]. Magnetization was measured using a Quantum Design SQUID magnetometer MPMS in the temperature range 5–700 K and magnetic field range 0–70 kOe. $^{45}$Sc NMR experiments were performed by the conventional spin-echo method using a Thamway PROT-II spectrometer. The sample was powdered and fixed with paraffin to ensure random orientation of the particles. The parameters of the $^{45}$Sc nucleus were nuclear spin $I = 7/2$, nuclear gyromagnetic ratio $\gamma = 1.0343$ MHz/kOe, nuclear quadrupole moment $Q = -0.220 \times 10^{-24}$ cm$^2$, and natural abundance ratio 100%. DFT calculations were performed using the Vienna ab initio simulation package (VASP) [19-22]. We used the projector augmented wave (PAW) pseudopotentials [23, 24] with the generalized gradient approximation (GGA) scheme following the Perdew, Burke and Ernzerhof (PBE) functional [25]. The conjugate gradient algorithm was used to relax the atoms. The Methfessele-Paxton scheme [26] was used for both geometry relaxation and total energy calculations. All atoms were relaxed until the forces on the atoms were less than of $10^{-2}$ eV/Å and the energy difference between two successive electronic steps was less than $10^{-7}$ eV. The unit cell was doubled along the $c$-axis and the $k$-point mesh was set to $35 \times 35 \times 5$ when spin-orbit coupling was not considered and $17 \times 17 \times 3$ when it was considered.

The inset of Fig. 2 shows examples of isothermal magnetization curves (5 and 300 K). They vary linearly through the origin, indicating no $M_s$ at the temperatures, contrary to what was reported in [18].

In our experiments, when the sample was synthesized with excess Fe, a small $M_s$ was observed. This is most likely due to FM impurities based on the hexagonal Laves phase $ScFe_2$ with $T_C$ = 542 K [27]. The temperature dependence of the susceptibility χ measured in a temperature range up to 700 K and a field of 10 kOe is shown in Fig. 2. χ is small at low temperatures, decreases slightly with increasing temperature up to ~400 K, increases at higher temperatures, reaches a maximum at ~650 K, and decreases at higher temperatures. Another small hump was observed at ~570 K (denoted by $T^*$). Except for the hump at $T^*$, this behavior is typical for AFM materials, suggesting that $ScFe_6Ge_4$ is AFM below $T_N \approx 650$ K.

The Mössbauer spectrum of our sample at RT (not shown) was the same as reported in [18] (see Fig. 7 in [18]); the spectrum was practically a single sextet component with $|H_{hf}(Fe)| \approx 191$ kOe. Assuming a standard hyperfine coupling for the 3d transition metal elements, $\mu_{Fe}$ is estimated to be ~1–2 $\mu_B$. The observed $|H_{hf}(Fe)|$ is nearly identical to the RT value for $Fe_3Sn_2$ (199–200 kOe) [28, 29], suggesting that the $\mu_{Fe}$ of $ScFe_6Ge_4$ is comparable to that of $Fe_3Sn_2$; the reported $M_s$ in $Fe_3Sn_2$ is ~2 $\mu_B$/Fe [16].

It is also interesting to note that the $T_N$ of $ScFe_6Ge_4$ ($\approx 650$ K) is almost the same as the $T_C$ of $Fe_3Sn_2$, indicating that the magnetic interaction strength in $ScFe_6Ge_4$ and $Fe_3Sn_2$ is almost the same and differs only in sign. The sign difference can be attributed to the different medium driving the inter-bilayer interaction: $ScGe_8$ in $ScFe_6Ge_4$ and $Sn_8$ in $Fe_3Sn_2$ (see Figs. 1(a) and (b)). Note also that the $T_N$ of $ScFe_6Ge_6$ ($\approx 500$ K) in the absence of the bilayer blocks is significantly lower than that of $ScFe_6Ge_4$ ($\approx 650$ K), indicating that the intra-bilayer interaction is stronger than the inter-monolayer interaction. In $ScFe_6Ge_4$, the in-plane Fe-Fe bond length is 2.48 Å or 2.60 Å, and the inter-plane Fe-Fe bond length in the bilayer block is 2.97 Å, suggesting that the Fe-3d electrons hybridize directly in the bilayer block. On the other hand, the bilayer blocks are separated by more than 4 Å, suggesting an indirect magnetic coupling between the bilayer blocks. Considering that the same bilayer block is common to both FM and AFM materials, it is reasonable to assume that the magnetic coupling within the bilayer block is FM both in-plane and inter-plane.

To investigate the field dependence of the susceptibility humps observed at high temperatures, the temperature dependence of $M/H$ ($M$ is the magnetization) was measured at several different fields $H$ (inset of Fig. 3). Two humps corresponding to $T^*$ and $T_N$ were observed in the range of measured fields. $T_N$ changed little with field, while $T^*$ decreased slightly with increasing field. Considering that $T^*$ tentatively corresponds to a certain phase transition, a magnetic phase diagram is shown in Fig. 3.

There are two distinct AFM phases. The low-temperature phase appears to be slightly destabilized by the application of the field. The change in the spin structure of $Fe_3Sn_2$ pointed out by Fenner *et al*. [17] is suggestive; $Fe_3Sn_2$ undergoes a transition from the paramagnetic phase to a *c*-axis collinear FM state at $T_C$, but at 520 K the moments begin to tilt from the *c*-axis to the *c*-plane, *i.e.*, a non-trivial magnetic state with a non-collinear component begins to appear below the temperature close to $T^*$ of $ScFe_6Ge_4$. It is very likely that the same phenomenon occurs in $ScFe_6Ge_4$ as in $Fe_3Sn_2$.

The inset of Fig. 4 shows an example of a $^{45}$Sc-NMR field sweep spectrum at 4.2 K. A sharp spectrum with satellites was observed near the field corresponding to ν/γ, where ν is the experimental frequency. The resonance field, corresponding to the position of the center line of the spectrum, was measured at different frequencies and plotted in Fig. 4. Extrapolation of the data yields a straight line passing through the origin with a slope of 1.033(1) MHz/kOe, almost equal to γ of $^{45}$Sc, thus assigning the observed signal to the $^{45}$Sc resonance. We can also see that $H_{hf}$(Sc) is practically negligible. An electric field gradient with asymmetry parameter η = 0 is expected at the Sc site (3*a*) with site symmetry $\bar{3}m$; the satellites observed around the center line are due to quadrupole effects. The inset of Fig. 4 also shows the simulated result of the powder pattern assuming a small negative isotropic hyperfine shift $K_{iso}$ = –0.3% and a quadrupole frequency $ν_Q$ = 0.76 MHz. The agreement between experiment and calculation is good, proving that $H_{hf}$(Sc) ≈ 0. This is in contrast to |$H_{hf}$(Fe)| ≈ 191 kOe. These results provide microscopic and direct evidence that $ScFe_6Ge_4$ is not FM but AFM. Furthermore, one should consider a magnetic structure in which $H_{hf}$(Sc) is canceled out.

Based on the magnetic structures reported for $Fe_3Sn_2$ and $ScFe_6Ge_6$ with similar stacking in their crystal structures [2, 10, 11], we consider possible magnetic structures under the constraint $H_{hf}$(Sc) = 0. $Fe_3Sn_2$ is FM ($T_C$ ≈ 650 K) [16] and $ScFe_6Ge_6$ is AFM ($T_N$ ≈ 500 K) [12]. In both compounds, the in-plane Fe-Fe bonds are FM. For simplicity, only the magnetic couplings between the Fe kagome layers are considered here, *i.e.* only up/down degrees of freedom in the moment direction are considered. In considering the model for $ScFe_6Ge_4$, it is reasonable to impose the following constraints. (i) The in-plane Fe-Fe magnetic coupling is FM. (ii) The interplane coupling in the bilayer block is FM. (iii) The coupling between the bilayer blocks is AFM. Assumption (i) was made for commonality with $Fe_3Sn_2$ and $ScFe_6Ge_6$. The reason for assumption (ii) is discussed above. Assumption (iii) is based on the fact that $H_{hf}$(Sc) = 0. Since the hexagonal Fe atoms are symmetrically coordinated above and below the $ScGe_8$ bipyramid, if the Fe moments of the kagome layers have AFM coupling with respect to the *c*-plane containing the Sc atoms, the $H_{hf}$(Sc) (both isotropic and anisotropic) cancel at the Sc site in terms of symmetry. This is consistent with the fact that in $ScFe_6Ge_6$, the coupling between the Fe layers sandwiching the $ScGe_8$ bipyramids is AFM. The magnetic structure satisfying (i)–(iii) is uniquely determined and shown schematically in Fig. 1(b); the stacking of in-plane FM kagome layers is -UU-DD-UU-DD-UU-DD- (the magnetic unit cell is twice the size of the crystal unit cell, hereafter denoted -UU-DD-), where U and D denote the FM layers with up and down moments, respectively, and the dash (-) corresponds to the $ScGe_8$ bipyramidal layer. Note that the discussion here does not exclude the existence of a noncollinear component or the possibility of non-uniformity in $\mu_{Fe}$.

To verify the validity of the proposed AFM structure, we calculated the total energies of the spin-polarized states by DFT. First, we compared the total energies of the FM and AFM states without considering the spin-orbit coupling (SOC). In addition to the model proposed above, we also calculated the total energies for the -UD-DU-UD-DU-UD-DU- and -UD-UD-UD-UD-UD-UD- sequences. The results, listed in Table 1 as the difference from the value for the FM state, show that the AFM configurations are generally more stable than the FM one, and the -UU-DD- sequence proposed above is the most stable. Furthermore, we have calculated the total energies of the FM state and the proposed state by considering the SOC. The result is almost the same as that without the SOC (see Table 1). In this case, the stable solution is a structure with the magnetic moment parallel to the $c$-axis. The lattice constants of the AFM state obtained by structural optimization are $a$ = 5.0524 Å and $c$ = 20.0216 Å, which are close to the experimental values of $a$ = 5.071 Å and $c$ = 20.053 Å at RT. The calculated $\mu_{Fe}$ is 2.0 $\mu_B$. The experimental $\mu_{Fe}$ is currently unknown for $ScFe_6Ge_4$, but it agrees with the Mössbauer result. Our results are inconsistent with those in [18], but the reason is not clear since the assumed AFM structures is not explicitly stated there.

$ScFe_6Ge_4$ with a double-layered kagome lattice of Fe is shown for the first time to be an antiferromagnet with $T_N$ = 650 K. The absence of the internal field at the Sc site provides microscopic evidence for the AFM state. The analogy with the magnetic structures of $ScFe_6Ge_6$ and $Fe_3Sn_2$, which contain the same substructures as $ScFe_6Ge_4$, and $H_{hf}(Sc) \approx 0$, suggests FM coupling within the bilayer block both in-plane and inter-plane, and AFM coupling between the blocks. The validity of the structure was confirmed by DFT calculations. Since the Fe sublattice in $ScFe_6Ge_4$ is crystallographically equivalent to that in $Fe_3Sn_2$, $ScFe_6Ge_4$ is of interest as an AFM alternative to $Fe_3Sn_2$ (FM at $T_C$ = 650 K), a known topological magnetic material. We must await single crystal experiments to obtain an accurate phase diagram and neutron diffraction experiments to know the evolution of the microscopic magnetism.

The authors thank K. Midori for his help in the early stages of the study. This work was supported by JSPS KAKENHI Grant Numbers 22F22016, 22KF0206 and JST SPRING, Grant Number JPMJSP2110.

Table 1. The total energy difference of the AFM states measured with respect to the FM state, as estimated by DFT calculations (eV/unit cell).

| | Magnetic structure | | | |
| --- | --- | --- | --- | --- |
| | FM | AFM | | |
| | -UU-UU- | -UD-DU- | -UD-UD- | -UU-DD- |
| Without SOC | 0 | –0.020 | –0.429 | –0.687 |
| With SOC | 0 | – | – | –0.690 |

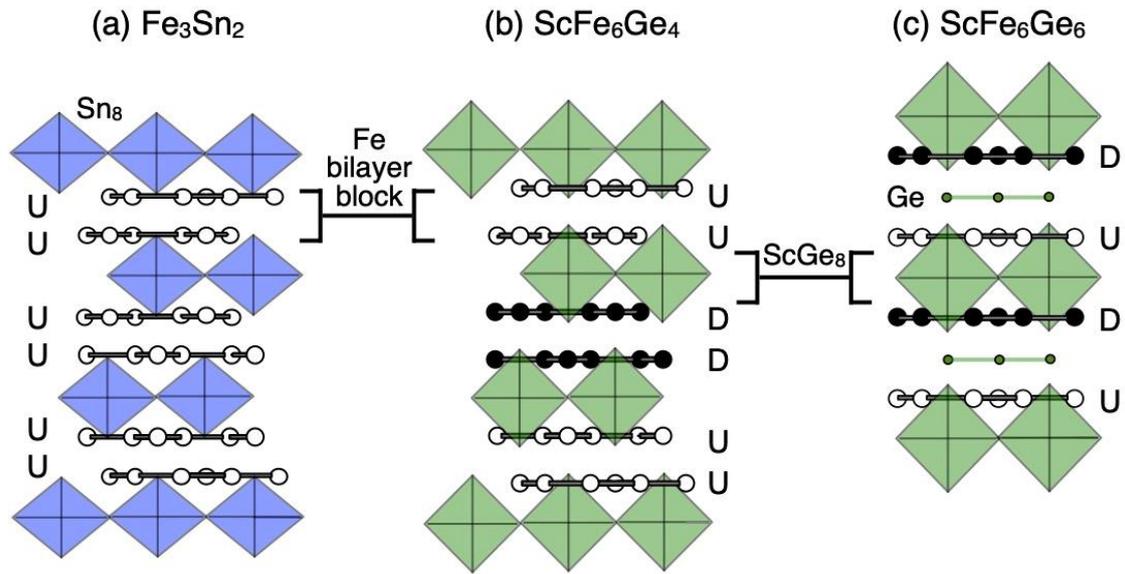

Figure 1. Interlayer magnetic couplings ([120] direction views) for $Fe_3Sn_2$ [16] (a) and $ScFe_6Ge_6$ [12] (c), and proposed magnetic coupling for $ScFe_6Ge_4$ (b). Open and closed circles represent Fe atoms (18$h$ site in space group $R\bar{3}m$) with up and down moments, respectively. Blue and green squares represent $Sn_8$ and $ScGe_8$ hexagonal bipyramids (dodecahedrons), respectively, with Sn/Ge and Sc atoms located at the corners and center, respectively. Small dots in (c) indicate the Ge honeycomb layer. In (b), only half of the magnetic unit cell is shown; the spin directions of adjacent parts in the $c$-direction are reversed.

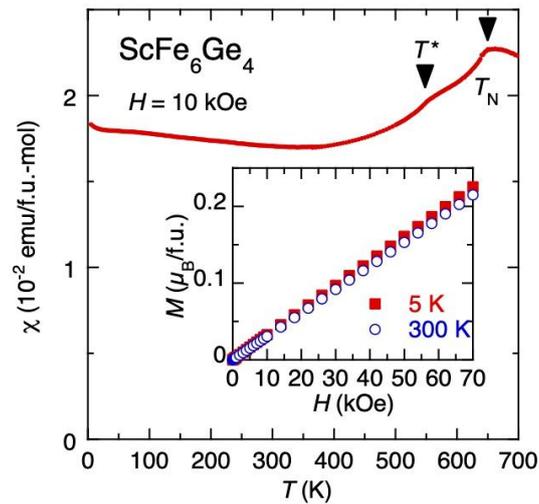

Figure 2. Temperature dependence of magnetic susceptibility measured at a magnetic field of 10 kOe for $ScFe_6Ge_4$. The inset shows examples of isothermal magnetization curves.

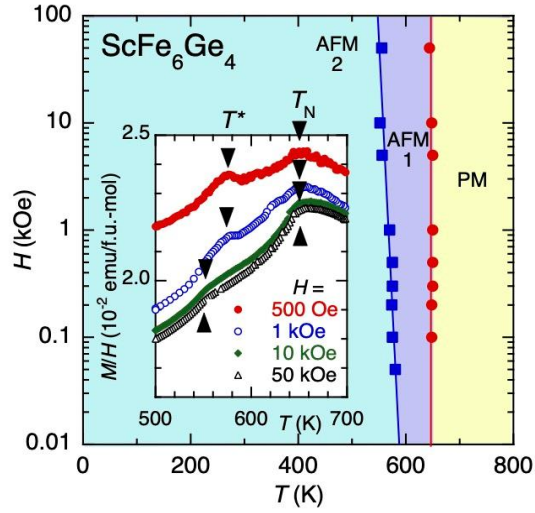

Figure 3. Magnetic phase diagram of ScFe$_6$Ge$_4$ evaluated from the susceptibility anomalies. The vertical axis is on a logarithmic scale. The inset shows temperature dependence of high-temperature part of magnetic susceptibility measured at different magnetic fields.

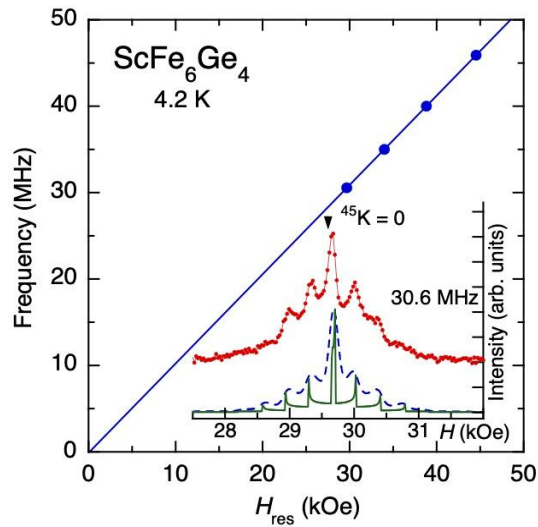

Figure 4. Frequency dependence of the resonance field of $^{45}$Sc NMR for ScFe$_6$Ge$_4$ at 4.2 K. The inset shows an example of the field-sweep spectrum (measurement frequency 30.6 MHz) (closed circles) and powder patterns calculated assuming $K_{iso} = -0.3\%$ and $\nu_Q = 0.76$ MHz (solid curve: bare intensity, dashed curve: Gaussian function with full width at half maximum of 150 Oe was convoluted).